\documentclass[manuscript,twocolumn]{aastex631}
\usepackage{float}
\usepackage{amsmath}  
\usepackage{graphicx}  

\revised{\today}
\tighten

\begin{document}
\shorttitle{Hypersaros and Similar Cycles} \shortauthors{Hartigan}

\title{A Search for Eclipse Cycles Similar to the Hypersaros: Columbus and the Lunar Eclipse of March 14, 2025

\vspace{0.45 in}}

\correspondingauthor{Patrick Hartigan} \email{hartigan@sparky.rice.edu}

\author[0000-0002-5380-549X]{Patrick Hartigan} \affiliation{Physics and Astronomy Dept., Rice University, 6100 S. Main, Houston,
TX 77005-1892}

\begin{abstract} 
\noindent
The total lunar eclipse on March 14, 2025 UT occurs nearly exactly 521 years (one Hypersaros) after a 
similar eclipse on March 1, 1504 UT that is renowned for its importance to the voyage of Columbus to 
Jamaica. Eclipses separated by a Hypersaros have similar depths, appear very close to the same 
location in the sky, and occur at nearly the same time of year. This paper summarizes the results from a 
search for analogous cycles within the Five Millennium Catalogs of Lunar and Solar Eclipses. Under the 
two simple constraints of similar eclipse dates relative to the vernal equinox and similar paths of the 
Moon through the Earth's shadow, the most common time intervals between lunar eclipses separated 
by less than 1000 years are the 521-year Hypersaros and a 633-yr period of the Icosa-Inex-Triple-Saros (IITS).  Notable cycles at longer periods occur at 1154, 1284, 1787, 1917, and 2308 years.

\end{abstract}


\section{Introduction} 
\label{sec:intro}

The grandeur and ease of observation of lunar and solar eclipses make them among the most
common astronomical phenomena experienced by the general public. When discussing these
events during outreach activities, astronomers often use the Saros (18.03 years) and, less commonly,
the Inex (28.95 years) series to link an upcoming eclipse with similar ones in the past and future. 
However, eclipses within a given Inex or Saros series do not occur at the same time of year,
and will appear higher or lower in the sky depending on the season even if the
longitude of the observer places the central eclipse on the meridian. 
From a purely observational point of view it makes sense to adopt a different approach
to group eclipses together. One possibility is to
take two eclipses to be related to each other
if they occur at nearly the same time of year and if the Moon traverses
almost an identical path across the Sun for a solar eclipse or through the Earth's shadow
for a lunar eclipse. Eclipses that occur at the same time of year are also located at the
same position in the celestial sphere.

As in the Inex series, the above prescription does not keep track of differing lunar distances.
The observational effect of differing lunar distances for lunar eclipses is minor,
limited to creating barely perceptible changes in the lunar angular diameter 
and in the duration of the eclipse (both effects $\lesssim$ $\pm$ 5\%). 
Grouping eclipses only by time of year and shadow path is less ideal for
solar eclipses, where the lunar distance determines whether a solar eclipse is
total, annular, or hybrid.  The above definition constrains the terrestrial latitude but not the
longitude of eclipse paths, and so does not guarantee that the Moon will be above
the horizon at any given location for lunar eclipses.
Variations in the length of the synodic month caused by the Moon's elliptical orbit 
\citep[e.g.][]{Meeus09}
introduce scatter of $\sim$ $\pm$ 12 hours in the UT of greatest eclipse.

\section{Cycle Identification and Analysis}
\label{sec:analysis}

This paper uses the Five Millennium Catalog of Lunar and Solar Eclipses \citep{EM09}
to identify pairs of eclipses that satisfy two criteria:
(1) {\it The difference $\Delta$T between the time separations of the two eclipses
from their respective vernal equinoxes must lie below some threshold.}
This requirement means that the two eclipses will occur at nearly
the same time of year, will be separated in time by nearly an
integral number of tropical years, and will have similar ecliptic longitudes. 
(2) {\it The difference $\Delta\gamma$
between the eclipse paths must be less than some limit, where $\gamma$ is
the distance of the shadow cone axis to the center of the Moon for a lunar
eclipse (to the center of the Earth for a solar eclipse) in units of
Earth's equatorial radii at the instant of greatest eclipse} \citep{EM09}. 
This requirement ensures that the path of the Moon through the
Earth's shadow (or its path across the Sun for solar eclipses) is
similar for the eclipse pair.  Gamma distinguishes between eclipses
where the Moon passes above the ecliptic plane during the eclipse
(positive $\gamma$) from those where the Moon moves 
below the ecliptic plane (negative $\gamma$).

Defined this way, the most common time separations between related eclipse pairs represent
eclipse cycles that connect directly with two easily observed parameters. 
Each pair of eclipses
that satisfies the $\Delta$T and $\Delta\gamma$ constraints contributes to the histogram for
a cycle with a period equal to the number of years between those eclipses.
By construction, the length of
these cycles will always be close to an integer number of years because related pairs of eclipses
have nearly the same calendar date.
If a cycle of T-years is 100\%\ reliable,
then every eclipse will have a counterpart of similar depth T years into the future and T years into the past.
Unlike the Inex and Saros series, with this definition a given eclipse will in general be a
part of several different time series, one for each match in date and depth that
occurs in the eclipse record.
There are over 140 million possible pairs of both lunar and solar eclipses in the Millennium catalogs.

Figure~1 displays the results of this exercise for two lunar eclipse examples, one with tight
constraints on the time interval and the path (Case A: $\Delta$T = $\pm$5~days
$\Delta\gamma$ = 0.2) and another with looser constraints (Case B:
$\Delta$T = $\pm$10~days and $\Delta\gamma$ = 0.3).  Prominent
peaks are marked with their periods in tropical years, their name if one exists from the 
compilation of R.H. van Gent\footnote{https://webspace.science.uu.nl/$\sim$gent0113/eclipse/eclipsecycles.htm
 2024, "A Catalogue of Eclipse Cycles,
Solar- and Lunar-Eclipse Predictions from Antiquity to the Present"}.
Solid lines denote the percentage of matches possible in the catalog for a given
cycle period. This number
declines with period because of the finite time interval of the catalog. If a peak 
in Fig.~1 at period P were to reach the 100\%\ line, then all of the 
eclipses in the catalog would have a corresponding eclipse P years later
that satisfies the $\Delta$T and $\Delta\gamma$ constraints. The looser constraints
of case B allow for about three times as many matches than for case A.
Approximately 0.12\%\ of the pairs satisfy Case A, and 0.36\%\ of the pairs satisfy Case B.
In practice, no cycle can ever attain 100\%\ because 
the Moon gradually drifts away from the ecliptic plane for all cycles until eventually eclipses
end, implying that eclipses within P-years of the end of a cycle will have no matches for
periods longer than P.

Both case A and case B have nearly 100\%\ cycles at the 521-year period of the
Hypersaros \citep[521.0107 years;][]{pogo35}, and at a 633-year 
period of 20 Inex plus 3 Saros that I refer to as the Icosa-Inex-Triple-Saros or IITS (632.9908 years).
Hence, almost every eclipse in the catalog is followed in 521 years and also in 633 years by
nearly identical eclipses.
Other strong periods with $\gtrsim$ 67\%\ matches for Case A include cycles at 130 (double McNaughton),
1154 (van den Bergh), 1787 (Heliotrope), 2308 (double van den Bergh) years, and unnamed cycles at
1284 [I], and 1917 [II] years, all of which are combinations of the Hypersaros, IITS, and the double
McNaughton cycles.  Results for case B show many discrete cycles, all various combinations
of three generative periods close to an exact integer number
of years: the 111.98-year Mercury, the 130.01-year double McNaughton, and an unnamed 279.02-year cycle.
Several weaker peaks also exist in both plots, including regular patterns at a level of $\sim$ 6\%\ for
Case A and 10\%\ for Case B that space out by the 19-year Metonic cycle.
The 65.005-year \citet{mcn95} cycle is not present in 
Fig.~1 because any odd number of Inex cycles changes the sign of $\gamma$; hence, eclipses
separated by 65 years typically do not satisfy the $\Delta\gamma$ criterion.
Corresponding graphs derived from the solar eclipse catalog are essentially identical to Fig.~1.

Given the strengths of the peaks at the Hypersaros and IITS, it is instructive to look at their
counterparts for an upcoming total lunar eclipse such as the one
visible throughout North America on March 14, 2025 UT. The past
Hypersaros counterpart for this event is the eclipse on March 1, 1504 UT (Julian calendar),
and the one in the future will be on March 18, 2546. The past eclipses for the Mercury and
double McNaughton cycles occurred on
March 22, 1913 and March 11, 1895, respectively, and the
past IITS eclipse was on March 9, 1392.  The 1504 eclipse is well-documented as
having been used by Columbus, who knew of the eclipse prediction,
to convince the native tribes in Jamaica to aid his crew.
Hence, observers of the 2025 eclipse have a chance to reprise that historical event.

\begin{figure*}[!t]
\centering
\includegraphics[width=0.7\linewidth]{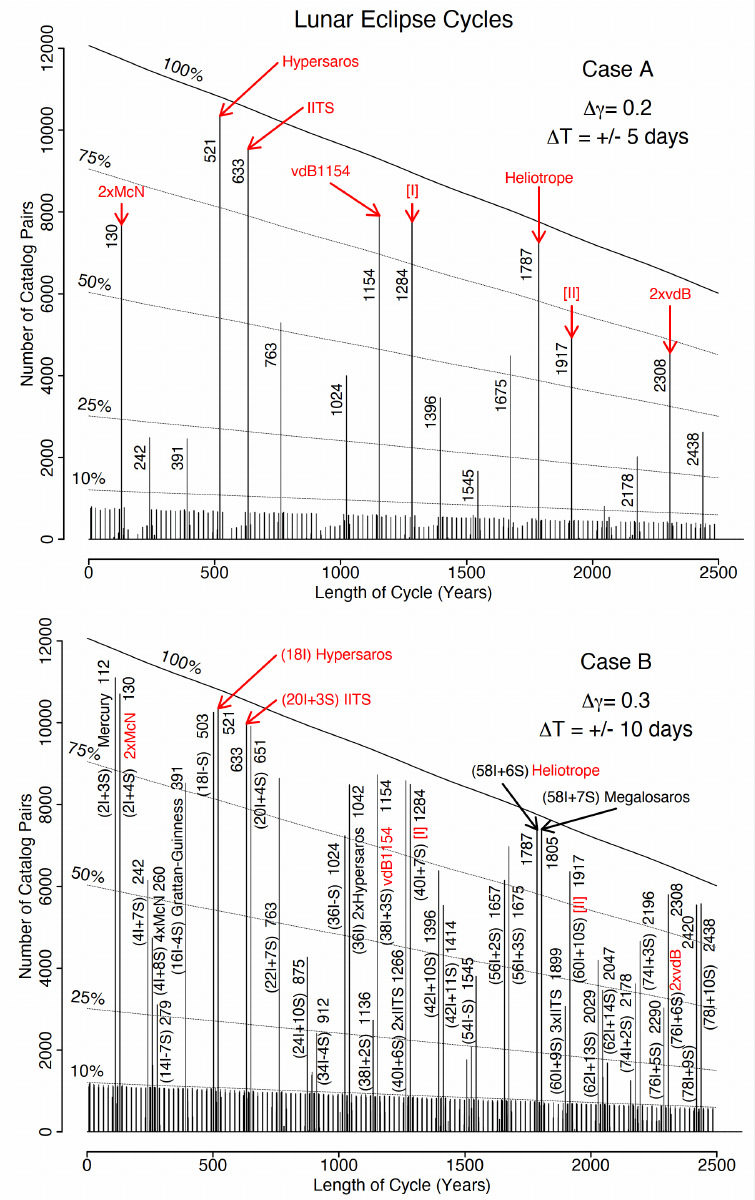}
\caption{
Lunar eclipse cycles defined by the criteria in the paper, with periods in years and 
Inex-Saros combinations in parentheses.  Black lines show fixed percentages 
of catalog matches for a given time delay.
Top: Case A, showing catalog matches for
$\Delta$T = $\pm$5~days, $\Delta\gamma$ = 0.2, 
Bottom: Case B, where $\Delta$T = $\pm$10~days, $\Delta\gamma$ = 0.3. 
The labeled peaks are all combinations of the 112-yr, 130-yr, and 279-yr cycles.
The Hypersaros peak (521 years; 18 Inex), the IITS peak (633 years; 20 Inex + 3 Saros)
are strong in both cases. Other notable peaks marked in red in both panels are the 2xMcN, vdB1154, Heliotrope, 2xvdB1154, and
Cycles I and II (130, 1154, 1787, 2308, 1284, and 1917 years, respectively).
}
\label{fig:cycle}
\end{figure*}
\hfill\break
This work relies entirely upon the eclipse compilations 
of Espanek and Meeus hosted at JPL. The author would like to thank
R. H. Gent for his compilation of the historical names for eclipse cycles.

\end{document}